\documentclass[aps,prb,twocolumn,superscriptaddress,reprint]{revtex4-2}
\usepackage{float}
\usepackage{graphicx}
\usepackage{amsmath}
\usepackage{bm}
\usepackage{color}
\usepackage{amssymb}
\usepackage{mathrsfs}
\usepackage{dcolumn}
\usepackage{multirow}
\usepackage{wasysym}
\usepackage[mathlines]{lineno}
\usepackage[colorlinks=true,linkcolor=blue,urlcolor=blue,citecolor=blue]{hyperref}

\begin{document}
	
	\title{Topological Aspects of Dirac Fermions in a Kagom\'e Lattice}

	\author{Xinyuan Zhou}
	\affiliation{Department of Physics, Zhejiang Normal University, Jinhua 321004, China}
	
	\author{Ziqiang Wang}
	\affiliation{Department of Physics, Boston College, Chestnut Hill, Massachusetts 02467, USA}
	
	\author{Hua Chen}
	\email{Electronic address: hwachanphy@zjnu.edu.cn} 
	\affiliation{Department of Physics, Zhejiang Normal University, Jinhua 321004, China}
	\affiliation{Zhejiang Institute of Photoelectronics \& Zhejiang Institute for Advanced Light Source, Zhejiang Normal University, Jinhua 321004, China}

	\begin{abstract}
		The Dirac fermion with linear dispersion in the kagom\'e lattice governs the low-energy physics of different valleys at two inequivalent corners of hexagonal Brillouin zone. The effective Hamiltonian based on the cyclic permutation symmetry of sublattices is constructed to show that the topology of Dirac fermions at these two valleys is characterized by opposite winding numbers. For spinless fermions, the many-particle interactions produce intervalley scattering and drive an intervalley coherent state with spontaneous translation symmetry breaking. The Dirac fermions acquire a mass from the simultaneous charge and bond orderings. In this phase, the developed bond texture underlies a hollow-star-of-David pattern in a tripled Wigner-Seitz cell of kagom\'e lattice. It is further demonstrated that the twisting of Dirac mass with vorticity leads to Dirac zero modes at the vortex core, which lead to fractionalization and anyon statistics. The hollow-star-of-David phase is shown to have a distinct $\mathbb{Z}_6$ Berry phase with its sign-change counterpart of Dirac mass, {\it i.e.} the hexagonal phase, shedding light on the topological origin of Dirac zero modes around the vortex core.
	\end{abstract}
	
	\date{\today}
	
	\maketitle
	
	\section{Introduction}
	The kagom\'e lattice with diverse quantum phenomena~\cite{Yin2022,Wang2023} holds promise to study the interplay between geometry, topology and correlation for its typical features in the band structure that includes Dirac cones~\cite{Nishimoto2010,OBrien2010,Rueegg2011,Ferhat2014,Ye2018,Kang2019,Yin2020,Liu2020,Peng2021,Hu2022}, quadratic band crossings~\cite{Liu2010,Wen2010,Zhu2016,Ren2018,Zeng2018}, van Hove singularities~\cite{Yu2012,Kiesel2012,Kiesel2013,Wang2013,Wu2021,Teng2022,Hu2022a,Kang2022,Ferrari2022,Liu2024a}, as well as flat bands~\cite{Lin2018,Kang2020,Li2021,Regmi2022,Sun2022,Hu2022b,Gao2023}. In particular, Dirac semimetals (DSMs) with linear dispersion are usually believed as a precursor of various topological insulating states~\cite{Shen2017,Bernevig2013}. The routine recipe for realizing topological insulators from DSMs is the inclusion of spin-orbit couplings, which can be traced back to the studies in graphene~\cite{Kane2005,Kane2005a}. In kagom\'e lattices, the intrinsic spin-orbit coupling preserving time-reversal symmetry opens a gap at Dirac points, giving rise to quantum spin Hall insulators with $\mathbb{Z}_2$ topology~\cite{Guo2009,Zhang2011,Bolens2019}. In addition, the ferromagnetic ordering breaking time-reversal symmetry further drives a staggered magnetic flux phase, rendering quantum anomalous Hall insulators with nontrivial Chern numbers~\cite{Xu2015,Yin2020}. 
	Besides topological insulators, another topic of topological relevance is the Dirac zero modes (DZMs) at vortex defects~\cite{Jackiw1976,Su1979,Jackiw1981,Read2000}, which are predicted in a Kekul\'e distortion of honeycomb lattices~\cite{Hou2007,Jackiw2007,Chamon2008} and lead to fractionalization and anyon statistics~\cite{Wilczek1982,Wilczek1983,Seradjeh2008,Greiter2024}. However, a comprehensive understanding of DZMs is still missing for kagom\'e lattices. In this work, we show that the many-particle interactions can drive the Dirac fermions to acquire a mass in a distorted kagom\'e lattice with the bond order of hollow-star-of-David (HSD) pattern. Notably, this bond and charge density wave order at the Dirac filling in a kagom\'e seimetal is complementary to the bond charge-density wave order with the star-of-David pattern at the van Hove filling in the kagom\'e metals~\cite{Jiang2021,Tan2021}. It is also shown that DZMs exist in the vortex excitations of HSD phase with a localized phase-winding in the mass of Dirac fermions. We further demonstrate that the HSD phase and its sign-change-of-mass counterpart hexagonal phase can be topologically distinguished by the $\mathbb{Z}_6$ Berry phase. Bound states at the interface between domains governed by distinct topological invariants are topologically protected, thereby revealing the topological origin of DZMs around the vortex core. 
	
	The rest of our study is organized as follows. In Sec.~\ref{sec:TBE}, we introduce the tight-binding model and derive the effective Hamiltonian to describe the Dirac fermions. Section~\ref{sec:MF} presents the detailed mean-field calculations. The topological properties are further discussed in Secs.~\ref{sec:DZM} and \ref{sec:Berry}. We summarize and discuss the results in Sec.~\ref{sec:summary}. 
	
	\begin{figure}
		\centering
		\includegraphics[width=0.5\textwidth]{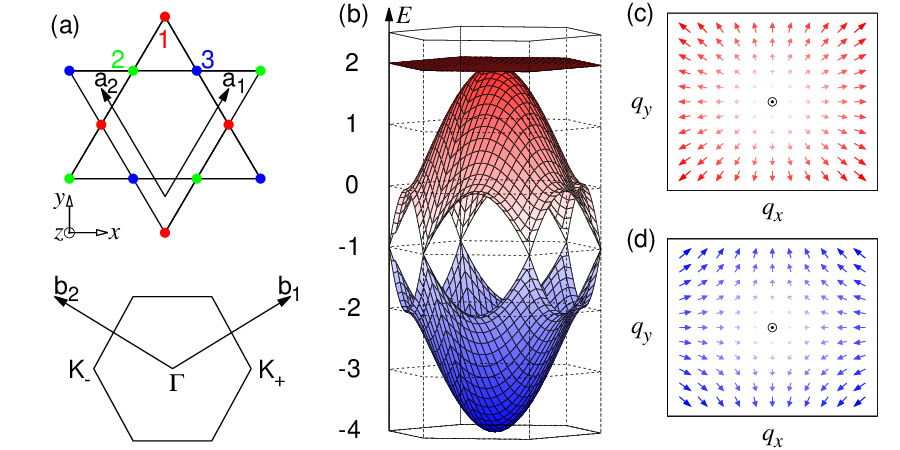}	
		\caption{
			(a) Kagom\'e lattice is a triangular Bravais lattice with three sublattices in a unit cell and the corresponding hexagonal Brillouin zone in reciprocal space. 
			(b) The band structure of tight-binding model in Eq.~(\ref{eq:TB}) with nearest-neighbor hopping $t=1$. 
			The pseudovector field $\bm{p}\equiv\left(p_x,p_y\right)$ near the Dirac point at the corner of hexagonal Brillouin zone $K_+$ (c) and $K_-$ (d) with the winding number $\mathcal{W}=\pm1$ in Eq.~(\ref{eq:wind}) resembles the vortex in $\text{XY}$ systems.
		}
		\label{fig:band}
	\end{figure}
	
	\section{Tight-binding model and effective Hamiltonian}
	\label{sec:TBE}
	We begin with the tight-binding model
	\begin{equation}
		H_{\text{TB}}=-t\sum_{\langle ij \rangle}c^\dagger_{i}c_{j}
		\label{eq:TB}
	\end{equation}
	that describes the nearest-neighbor hopping processes of spinless fermions in the kagom\'e lattice, as depicted in Fig.~\ref{fig:band}(a). Here, $t$ is the hopping integral, $c_{i}$ and $c^\dagger_{i}$ are the annihilation and creation field operators at $i$th site, respectively. Introducing a spinor representation of three sublattices $c_{\bm k}=\left[c_{1{\bm k}},c_{2{\bm k}},c_{3{\bm{k}}}\right]^\text{T}$, the Bloch Hamiltonian with the unit-cell gauge in reciprocal space reads
	\begin{equation}
		\mathcal{H}_{\bm{k}} =
		\hspace{-1.2mm}
		\begin{bmatrix}
			0						&
			1+\exp\left[ik_1\right]&
			1+\exp\left[ik_2\right]\\
			1+\exp\left[-ik_1\right]&
			0						&
			1+\exp\left[ik_3\right] \\
			1+\exp\left[-ik_2\right]&
			1+\exp\left[-ik_3\right]&
			0																			
		\end{bmatrix}
		\label{eq:hk}
	\end{equation}
	with the crystal momenta $k_{i=1,2,3}=\bm{k}\cdot\bm{a}_i$ being expressed in terms of the primitive lattice vectors $\{\bm{a}_1,\bm{a}_2,\bm{a}_3\equiv -\bm{a}_1+\bm{a}_2\}$ respectively. The band structure of the tight-binding model in Eq.~(\ref{eq:TB}) is plotted in Fig.~\ref{fig:band}(b). Notably, the highest band is a perfectly flat band with a quadratic band touching at $\Gamma$ point and the lowest two bands cross at the corners $K_\pm$ points of hexagonal Brillouin zone (HBZ). Moreover, the Bloch Hamiltonian $\mathcal{H}_{\bm k}$ in Eq.~(\ref{eq:hk}) at $K_\pm$ points is left invariant under the cyclic permutation of three sublattices. Therefore, the corresponding Bloch states are characterized by the irreducible representations of group $C_3$. To describe the corresponding low-energy behavior around $K_\pm$ points, we further derive the effective $k\cdot p$ model  
	\begin{eqnarray}
		\mathcal{H}_{K_\pm} \left({\bm q}\right) = 
		p_0\tau_0
		+p_x\tau_x
		+p_y\tau_y
		+\mathcal{O}\left(q^2\right)
		\label{eq:kp}
	\end{eqnarray}
	with the coefficients $p_0 = -t$ and $\{p_x,p_y\} = \sqrt{3}t\{\pm q_x,q_y\}$.
	Here, $\tau_0$ and $\bm{\tau}$ are the identity matrix and Pauli matrices, respectively, operating on the space spanned by the irreducible representations $E$ and $A$ of group $C_3$. 
	The detailed derivation is presented in Supplemental Material~\cite{SM}. 
	Diagonalizing $\mathcal{H}_{K_\pm}\left({\bm q}\right)$ in Eq.~(\ref{eq:kp}) yields two isotropic bands $E^\pm_{K_\pm}\left({\bm q}\right)=-t\pm\sqrt{3}t\left|\bm{q}\right|$, resulting in a linear dispersed Dirac cone with Fermi velocity $v_F=\sqrt{3}t/\hbar$. The pseudovector fields $\bm{p}\equiv\left(p_x,p_y\right)$ around $K_\pm$ points, shown in Figs.~\ref{fig:band}(c) and \ref{fig:band}(d) respectively, have a $p$-wave symmetry. The topological charge of Dirac point is given by the winding number of pseudovector field
	\begin{equation}
		\mathcal{W}=\frac{1}{2\pi}\oint_\mathcal{C}\nabla\varphi\left({\bm q}\right)\cdot d{\bm q}=\pm1
		\label{eq:wind}
	\end{equation}
	for the $K_\pm$ point respectively, where $\varphi\equiv\text{arctan}\left(p_y/p_x\right)$ and $\mathcal{C}$ is a contour enclosing the singular $K_\pm$ points, indicating that the Dirac point carries a $\pm\pi$ Berry flux. 
	
	\begin{figure}
		\centering
		\includegraphics[width=0.5\textwidth]{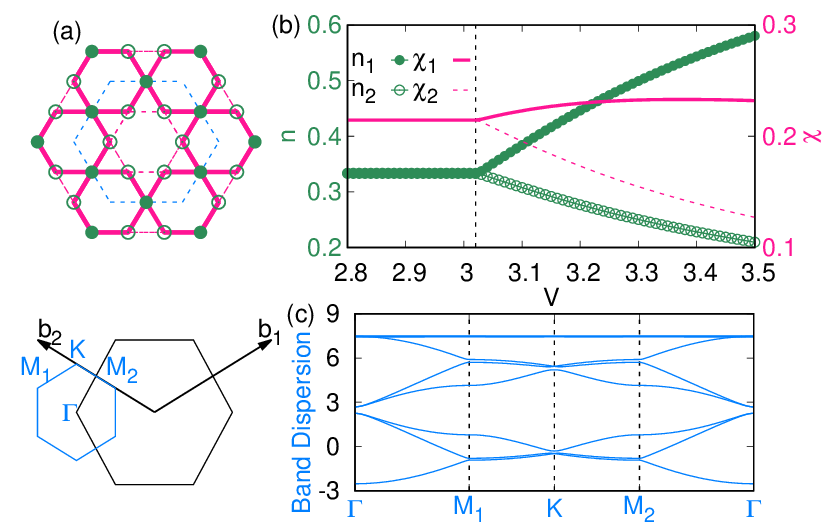}	
		\caption{
			(a) The pictorial representation of charge and bond patterns for the hollow-star-of-David insulator phase in an enlarged Wigner-Seitz unit cell indicated by blue dash lines, and the corresponding reduced (original) hexagonal Brillouin zone indicated by blue (black) lines in reciprocal space. 
			(b) The evolution of self-consistent mean-field order parameters including the charge density $n$ with filled (open) circles indicating large (small) local on-site charge, as well as the bond order $\chi$ with solid (dash) lines indicating strengthened (weakened) inter-site bond hopping renormalizations, as a function of inter-site interactions $V$ with near-neighbor hopping $t=1$ at the Dirac filling $\nu=1/3$.
			The black dashed line at $V_\text{c}\approx3.02$ marks the phase transition between the Dirac semimetal and hollow-star-of-David insulator phases with its charge and bond patterns in (a).
			(c) The band structure along high-symmetry paths in Brillouin zone with $\{t,V\}=\{1,3.1\}$. 
		}
		\label{fig:hf}
	\end{figure}
	
	\section{Mean-field calculations}
	\label{sec:MF}
	Having established the single-particle physics, we are then in a position to study the effects of many-particle interactions. The interacting Hamiltonian of spinless fermions described by the inter-site interaction can be decoupled at the Hartree-Fock mean-field level
	\begin{eqnarray}
		H_\text{I}=V\sum_{\langle ij \rangle}\left(n_i c^\dagger_jc_j+n_jc^\dagger_ic_i-n_in_j \right.\nonumber\\
		\left.-\chi_{ij}c^\dagger_jc_i-\chi_{ji}c^\dagger_ic_j+\left|\chi_{ij}\right|^2\right),
		\label{eq:hint}
	\end{eqnarray}
	where $n_{i}=\langle c_{i}^\dagger c_{i}\rangle$ and $\chi_{ij}=\langle c^\dagger_ic_j\rangle$ are the charge and bond mean-field order parameters, respectively. The charge $n$ related terms in Eq.~(\ref{eq:hint}) promote charge density waves, and the bond $\chi$ related ones renormalize the hopping integral $t$ in tight-binding Hamiltonian Eq.~(\ref{eq:TB}) by lowering the ground-state energy. Before proceeding, it is instructive to discuss the many-particle instability from the weak-coupling limit with the Fermi energy being pinned at Dirac points. The single-particle Bloch states around the corners $K_\pm$ points of HBZ are characterized by Dirac fermions at two inequivalent valleys. The many-particle interactions may scatter the Dirac fermions via the intra- and inter-valley processes. For the latter, the umklapp scattering between the corners of HBZ transfers a lattice phonon which carries the momentum of multiple primitive reciprocal vectors. This process folds two $K_\pm$ points to $\Gamma$ point and underlies the reduced HBZ (RHBZ), accompanied by the tripling of Wigner-Seitz cell in Kagom\'e lattices, as illustrated by the blue dash lines in Fig.~\ref{fig:band}(a). The $\sqrt{3}\times\sqrt{3}$ ordered states are theoretically predicted at first~\cite{Nishimoto2010} and experimentally observed in vanadium based kagom\'e metal ScV$_6$Sn$_6$~\cite{Arachchige2022,Hu2023a,Lee2024}. Besdies, similar intervalley coherent states have been extensively studied in the context of twisted bilayer graphene ~\cite{Kang2019a,Seo2019,Bultinck2020,Bernevig2021}. In our numerical calculations, the order parameters in both the charge $n$ and bond $\chi$ channels are self-consistently determined. Figure~\ref{fig:hf}(b) shows the evolution of order parameters as a function of inter-site interaction $V$ with hopping integral $t=1$ at the Dirac filling $\nu=1/3$. A critical interaction $V_\text{c}\approx3.02$ is found to separate DSM phase and hollow-star-of-David insulator (HSDI) phase. The HSDI phase, which is essentially driven by the intervalley scattering process, spontaneously breaks the translation symmetry of kagom\'e lattice with its charge and bond patterns sketched in Fig.~\ref{fig:hf}(a). The strengthened bonds renormalized by real-valued order parameters $\chi$ in an enlarged Wigner-Seitz unit cell highlight a HSD pattern with its vertices being composed of large on-site charge $n$. In contrast, the complex-valued bond orders at the van Hove fillings break time reversal symmetry with spontaneous loop currents~\cite{Feng2021,Dong2023}. To facilitate our understanding on the low-energy physics, we next derive the effective Hamiltonian in the vicinity of Dirac points $K_\pm$ by projecting the mean-field Hamiltonian in Eq.~(\ref{eq:hint}) onto the irreducible representations of group $C_3$ through a perturbative treatment. 
	The detailed derivation is presented in Supplemental Material~\cite{SM}. 
	Introducing a pseudospin $\sigma_z=\pm1$ to label the valley degree of freedom at $K_\pm$ points, the resulting Hamiltonian takes the following form
	\begin{equation}
		\mathcal{H}_\text{I} = \sqrt{3}\delta t\left(q_x\tau_x\sigma_z+q_y\tau_y\right)
		+ \Delta \tau_x\sigma_x
		\label{eq:hieff}
	\end{equation}
	with $\delta t=\frac{V}{2}\left(\chi_+ + \frac{\chi_-}{3}\right)$ and $\Delta=\frac{V}{3}\left(n_- - 2\chi_-\right)$ being expressed in terms of charge $n_\pm=n_1\pm n_2$ and bond $\chi_\pm=\chi_1\pm\chi_2$ order parameters. In Eq.~(\ref{eq:hieff}), we have dropped the renormalization of Fermi energy, which is irrelevant to the low-energy physics. The bond orders of HSD pattern in the first term of Eq.~(\ref{eq:hieff}) equally renormalize the Fermi velocity $v_F=\sqrt{3}\left(t+\delta t\right)/\hbar$ of Dirac fermions at $K_\pm$ points. Interestingly, the Dirac fermions, acquiring a mass term described by the charge and bond orders in the second term of Eq.~(\ref{eq:hieff}), become gapped by the intervalley scattering. The band structure along high-symmetry paths of RHBZ in Fig.~\ref{fig:hf}(c) is numerically calculated to confirm the insulating gap at the center $\Gamma$ point of RHBZ in HSDI phase. 
	
	\begin{figure}
		\centering
		\includegraphics[width=0.5\textwidth]{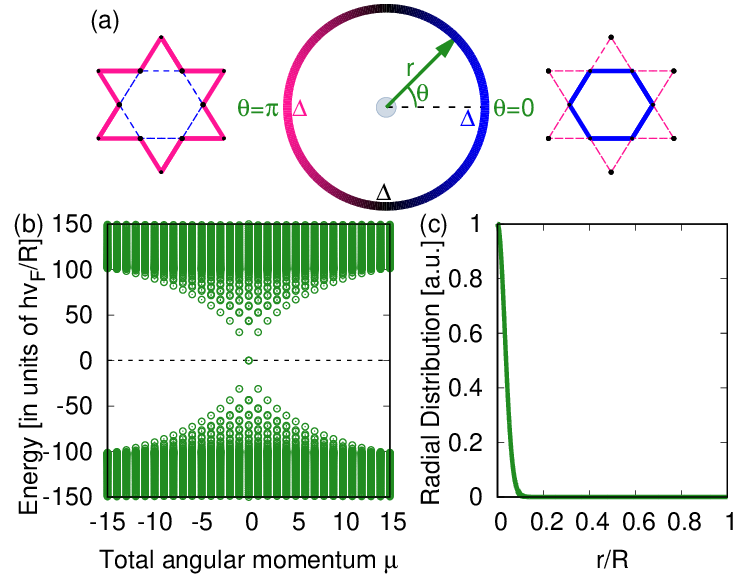}	
		\caption{
			(a) Schematic picture for the bond textures at polar angle $\theta=0$ and $\pi$ around the vortex with vorticity $\ell=-1$.
			(b) The energy spectrum of Dirac fermions immersed in a vortex $\Delta_{\ell=-1}\left(\bm{r}\right)=\Delta\left(r\right)\exp\left[-i\theta\right]$ with $\Delta\left(r\right)=\Delta\tanh\left(r/\xi\right)$ as a function of the eigenvalue $\mu$ of total angular momentum operator in Eq.~(\ref{eq:jz}). The strength and size of vortex core are set as $\Delta=100\hbar v_F/R$ and $\xi=R/5$, respectively. In practical implement of numerical calculations, the number of zeros of the Bessel function $N=500$ is chosen to ensure the convergence of the low-energy spectrum.
			(c) The radial distribution of the Dirac zero mode localized at vortex core with total angular momentum $\mu=0$.
		}
		\label{fig:vortex}
	\end{figure}
	
	\section{Dirac zero modes}
	\label{sec:DZM}
	Having settled the nature of HSDI phase, we then study DZMs in the presence of a vortex, which is intimately related to fractionalization and anyon statistics as its topological manifestation. 
	To this end, we generalize the mass term $\Delta$ in Eq.~(\ref*{eq:hieff}) with spatial dependence by focusing on the following Hamiltonian 
	\begin{equation}
		\mathcal{H}_\text{V}= \hbar v_F \left(\hat{q}_x\tau_x\sigma_z+\hat{q}_y\tau_y\right)
		+ \left(\Delta_\ell\left(\bm{r}\right)\sigma_+ 
		+ \Delta^*_\ell\left(\bm{r}\right)\sigma_-\right)\tau_x,
		\label{eq:hvor}
	\end{equation}
	where the vortex potential $\Delta_\ell\left(\bm{r}\right)=\Delta\left(r\right)\exp\left[i\ell\theta\right]$ in polar coordinates $\bm{r}=\left(r,\theta\right)$ has vorticity $\ell$ and Pauli matrices $\sigma_\pm=\frac{1}{2}\left(\sigma_x\pm i\sigma_y\right)$. Generally, the bond orders in the mass term $\Delta$ of Eq.~(\ref{eq:hieff}) develop a complex phase winding as the complex-valued Dirac mass around a vortex core~\cite{Wen2010,Rueegg2011}. Figure~\ref{fig:vortex}(a) shows the bond textures of hexagonal and HSD patterns at polar angle $\theta=0$ and $\pi$ modulated by the vortex potential $\Delta_{\ell=-1}\left(\bm{r}\right)$, respectively. It is easy to verify that the Hamiltonian in Eq.~(\ref{eq:hvor}) commutes with the following total angular momentum operator
	\begin{equation}
		J_z=L_z+\frac{\hbar}{2}\sigma_z\left(\tau_z-\ell\right)
		\label{eq:jz}
	\end{equation}
	where $L_z=-i\hbar\partial_\theta$ is the orbital angular momentum. Consequently, the Hamiltonian shares the same eigenstates with the total angular momentum operator. The eigenstates for polar angle $\theta$ can be labeled by the eigenvalue $\mu$ of operator $J_z$. The radial eigenstates of the Hamiltonian can be numerically solved through the Bessel-Fourier transformation on a disk geometry with the principal quantum number $E$ labeling the energy level. This numerical method is widely used to solve the Bogoliubov-de Gennes equation ever since the study of superconductivity~\cite{Gygi1991}. The Hilbert space is spanned by a set of renormalized orthogonal bases
	\begin{equation}
		\phi_{j,\ell}\left(r\right)=\frac{\sqrt{2}}{RJ_{\ell+1}\left(x_{j\ell}\right)}
		J_\ell\left(x_{j\ell}\frac{r}{R}\right)
	\end{equation}
	with $x_{j\ell}$ denoting the $j$th zero of the Bessel function $J_\ell\left(x\right)$, which imposes the Dirichlet boundary condition at maximum radius $R$ of the disk geometry. Expanding the space of valley pseudospin $\bm{\sigma}$, the Hamiltonian in subspace labeled by the eigenvalue $\mu$ of operator $J_z$ reduces to the following matrix
	\begin{equation}
		\mathcal{H}^\mu_{\text{V}} =
		\begin{bmatrix}
			\hbar v_Fq_-	&
			\Delta_+		\\
			\Delta_-		&
			\hbar v_Fq_+																			
		\end{bmatrix} _\sigma
		\tau_+ 
		+ 
		\begin{bmatrix}
			\hbar v_Fq_-^\dagger	&
			\Delta_-^\dagger		\\
			\Delta_+^\dagger		&
			\hbar v_Fq_+^\dagger																			
		\end{bmatrix} _\sigma
		\tau_-
	\end{equation}
	where Pauli matrix $\tau_\pm=\frac{1}{2}\left(\tau_x\pm i\tau_y\right)$, the corresponding matrix elements
	$q_\pm^{jj^\prime}=\int dr r\phi_{j,\mu\mp\frac{\ell-1}{2}}\hat{q}_\pm\phi_{j^\prime,\mu\mp\frac{\ell+1}{2}}$, 
	and $\Delta_\pm^{jj^\prime}=\int dr  r\phi_{j,\mu\pm\frac{\ell-1}{2}}\Delta\left(r\right)\phi_{j^\prime,\mu\mp\frac{\ell+1}{2}}$. The calculated energy spectrum, shown in Fig.~\ref{fig:vortex}(b), is obtained via the diagonalization of the Hamiltonian matrix. In numerical calculations, the vortex is described by $\Delta_{\ell=-1}\left(\bm{r}\right)=\Delta\tanh\left(r/\xi\right)\exp\left[-i\theta\right]$ with $\Delta$ and $\xi$ characterizing the strength and size of vortex core, respectively. The spectrum above the gap $\left|E\right|>\Delta$ is essentially continuous with the states being described by plane waves, but is actually discrete due to the quantum confinement of a finite disk. In contrast, the states below the gap $\left|E\right|<\Delta$ are bound states in nature with a discrete spectrum, for which the Dirac fermions are confined by its mass term~\cite{DeMartino2007}. Of most significance are the bound midgap states with total angular momentum $\mu=0$ at zero energy. The DZMs can be further identified through the analytical solutions of the Schr\"{o}dinger equation with its Hamiltonian in Eq.~(\ref{eq:hvor}). The linearly independent wave functions of these states up to an overall renormalization factor for generic vortex potentials are described by
	\begin{equation}
		\left|\pm\right>=e^{\pm\frac{1}{\hbar v_F}\int^r\Delta\left(r^\prime\right)dr^\prime}
		\left|\bm{\tau}=-\hat{z},\bm{\sigma}=\pm \hat{y}\right>
		\label{eq:zero}
	\end{equation}
	where Dirac notion $\left|\bm{\tau},\bm{\sigma}\right>$ is the tensor product of pseudospin space. The DZMs for vorticity $\ell=+1$ are given by reversing the pseudospin direction $\bm{\tau}=+\hat{z}$ in Eq.~(\ref{eq:zero}). The DZM $\left|+\right>$ bounded near $r=R$ can be regard as the chiral edge state of topological insulators with time-reversal symmetry breaking by the vortex. In contrast, the other DZM $\left|-\right>$ is bounded at the vortex core with its mechanism explored below. Unfortunately, direct comparisons between analytical and numerical solutions are unavailable because of different boundary conditions. Numerically, the DZMs can be distinguished on application of a radial electric field by lifting the degeneracy. Figure~\ref{fig:vortex}(c) plots the radial distribution of the DZM localized at the vortex core. Interestingly, a unitary transformation $U=\left(\sigma_++\sigma_-\tau_x\right)\sigma_x$ on the Hamiltonian in Eq.~(\ref{eq:hvor}) yields an effective Hamiltonian describing the proximity induced superconductivity on the surface of topological insulators, which supports a single Majorana bound state at the vortex core~\cite{Fu2008}.
	
	\begin{figure}
		\centering
		\includegraphics[width=0.5\textwidth]{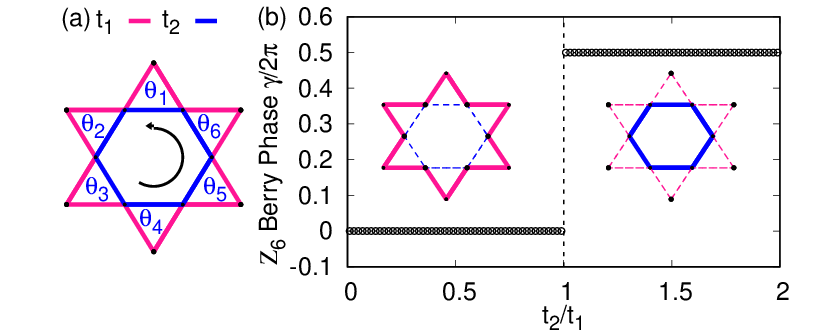}	
		\caption{
				(a) Schematic picture for the local twist phase on one chosen hexagon.
				(b) The $\mathbb{Z}_6$ Berry phase $\gamma$ in Eq.~(\ref{eq:berry}) as a function of the ratio of hopping integrals $t_2/t_1$. The bond textures (insert) for $t_2/t_1<1$ and $t_2/t_1>1$ regimes correspond the hollow-star-of-David and hexagonal patterns, respectively. These patterns are separated by the critical phase of Dirac semimetals  marked by the dash line at $t_1=t_2$. The numerical calculation is performed on a 1323-site cluster of hexagonal shape with the twist phase being introduced on one chosen hexagon. 
		}
		\label{fig:berry}
	\end{figure}
	
	\section{$\mathbb{Z}_6$ Berry phase}
	\label{sec:Berry}
	The effective low-energy Hamiltonian in Eq.~(\ref{eq:hvor}) is universal in describing the physics of zero-energy modes at the vortex core. In order to precisely pin down the topological origin of DZMs, we further investigate $\mathbb{Z}_6$ Berry phase with the tight-binding model, which is unique to the kagom\'e lattice. To proceed, a local twist phase, as shown in Fig.~\ref{fig:berry}(a), is introduced  on one chosen hexagon in a large cluster under the constraint $\sum_{i=1}^{6}\theta_i=0$. The parameter space can be expressed as $\bm{\Theta}=\sum_{i=1}^{5}\theta_i\bm{f}_i$ in terms of five independent phases $\theta_i$ with $\bm{f}_i$ being orthogonal unit bases. The $\mathbb{Z}_6$ Berry phase is then defined as a contour integral
	\begin{equation}
		\gamma=\int_{L_i} \bm{\mathcal{A}}\left(\bm{\Theta}\right)\cdot d\bm{\Theta}\text{ mod }2\pi
		\label{eq:berry}
	\end{equation}
	with $\bm{\mathcal{A}}\left(\bm{\Theta}\right)=-i\langle\psi\left(\bm{\Theta}\right)\left|\nabla_{\bm{\Theta}}\right|\psi\left(\bm{\Theta}\right)\rangle$ being the Berry connection of the ground state with twist phases $\bm{\Theta}$~\cite{Hatsugai2011}. The path $L_i$ in parameter space $\bm{\Theta}$ is defined as $\bm{E}_{i-1}\rightarrow\bm{G}\rightarrow\bm{E}_{i}$ with $\bm{E}_{i=1-5}=2\pi\bm{f}_i$, $\bm{E}_6=\bm{E}_0=\bm{0}$, and $\bm{G}=1/6\sum_{i=1}^{6}\bm{E}_i$. Generally, $\mathbb{Z}_6$ Berry phase, equivalent between different paths $L_i$, is quantized as $\gamma=2\pi\mathbb{Z}/6$ with $\mathbb{Z}$ being an integer. The details are presented in Supplemental Material~\cite{SM}. Figure~\ref{fig:berry}(b) shows that $\mathbb{Z}_6$ Berry phase $\gamma=0$ and $\pi$ for HSD ($t_1>t_2$) and hexagoanl ($t_1<t_2$) bond textures, respectively. The HSD ($\Delta<0$) and hexagonal ($\Delta>0$) insulating phases with the opposite masses, c.f. $\Delta$ in Eq.~(\ref{eq:hieff}), are separated by the critical phase of DSMs at $t_1=t_2$. Next, we turn to discuss the topological protection of DZMs in analogy with the Jackiw-Rebbi soliton at domain walls in the Su-Schriefer-Heeger model~\cite{Jackiw1976,Su1979,Su1980,Niemi1986}. Recently, the Jackiw-Rebbi soliton successfully applies in explaining the corner states~\cite{Xue2018,Li2019,ElHassan2019,Kempkes2019} of second-order topological insulators~\cite{Benalcazar2017,Benalcazar2017a,Song2017,Schindler2018a,Ezawa2018,Lu2020,Sil2020,Herrera2022,Zhou2023}. Figure~\ref{fig:vortex}(a) illustrates the domains with $\theta$-dependent masses $\Delta$ around the vortex core with vorticity $\ell=-1$. As polar angle $\theta$ evolves from $0$ to $2\pi$, the vertex core that connects different domains, serving as the domain wall, must experience a gap close and reopen process, due to distinct $\mathbb{Z}_6$ Berry phases at polar angle $\theta=0$ and $\pi$. It is indicative of in-gap states crossing the Fermi energy, corresponding to DZMs at the vortex core. Generically, such gap evolutions are involved for $\left|\ell\right|$ times as the polar angle changes from $\theta=0$ to $2\pi$ for vortex potential $\Delta_{\ell}\left(\bm{r}\right)$ in Eq.~(\ref{eq:hvor}), resulting the topologically protected DZMs of total number $\left|\ell\right|$.

	\section{Summary and discussion} 
	\label{sec:summary}
	To summarize, we have shown that the Dirac fermions in the semimetal state on the kagom\'e lattice become massive in a correlation driven bond and charge density ordered state with a HSD pattern. The existence of DZM is further numerically confirmed, arising from the interaction with vortex potential due to the vorticity and sharing the similar mechanism with the solitons at the domain wall of Dirac mass. Finally, its topological protection is also explained by $\mathbb{Z}_6$ Berry phase. Brief discussions on the related experiments are in order. The DZMs in a Kekul\'e distorted honeycomb lattice are experimentally observed in sonic, photonic and solid-state materials\cite{Gao2019,Menssen2020,Guan2024}, which are hopefully generalized to a distorted kagom\'e lattice of HSD pattern. Moreover, ultracold atoms also provide a promising platform~\cite{Bloch2008,Bloch2012,Gross2017} for the realization of our theoretical model by loading the fermionic atoms with a single hyperfine state into the optical kagom\'e lattice~\cite{Jo2012,Yamamoto2013}. Our findings may inspire the experimental searching for the topological phases of Dirac fermions in correlated kagom\'e materials.
	
	\section*{Acknowlegdgement}
	HC is supported by Zhejiang Provincial Natural Science Foundation of China under Grant No. LZ22A040002, and National Natural Science Foundation of China under Grants No. 12174345. ZW is supported by the U.S. Department of Energy, Basic Energy Sciences, Grant No. DE FG02-99ER45747.

	
	%

\end{document}